\documentclass[aps,prb,twocolumn]{revtex4-1}
\usepackage{amssymb}
\usepackage{graphicx}
\usepackage{hyperref}

\begin{document}
\author{Jaskaran Singh$^1$, Anooja Jayaraj$^1$, D. Srivastava$^1$, S. Gayen$^1$, A. Thamizhavel$^2$ and Yogesh Singh$^1$ }
\affiliation{$^1$Department of Physical Sciences, Indian Institute of Science Education and Research (IISER) Mohali, Knowledge City, Sector 81, Mohali 140306, India.\\
$^2$ Department of Condensed Matter Physics and Material Sciences, Tata Institute of Fundamental Research, Mumbai 400005, India.}

\date{\today}

\title{Possible multigap Type-I superconductivity in the layered Boride RuB$_2$}

\begin{abstract}
The structure of the layered transition-metal Borides $A$B$_2$ ($A =$ Os, Ru) is built up by alternating $T$ and B layers with the B layers forming a puckered honeycomb.  Here we report superconducting properties of RuB$_2$ with a $T_c \approx 1.5$~K using measurements of the magnetic susceptibility versus temperature $T$, magnetization $M$ versus magnetic field $H$, resistivity versus $T$, and heat capacity versus $T$ at various $H$.  We observe a reduced heat capacity anomaly at $T_c$ given by $\Delta C/\gamma T_c \approx 1.1$ suggesting multi-gap superconductivity.  Strong support for this is obtained by the successful fitting of the electronic specific heat data to a two-gap model with gap values $\Delta_1/k_BT_c \approx 1.88$ and $\Delta_2/k_BT_c \approx 1.13$.  Additionally, $M$ versus $H$ measurements reveal a behaviour consistent with Type-I superconductivity.  This is confirmed by estimates of the Ginzburg-Landau parameter $\kappa \approx 0.1$--$0.66$.  These results strongly suggest multi-gap Type-I superconductivity in RuB$_2$.  We also calculate the band structure and obtain the Fermi surface for RuB$_2$.  The Fermi surface consists of one quasi-two-dimensional sheet and two nested ellipsoidal sheets very similar to OsB$_2$.  An additional small $4^{\rm th}$ sheet is also found for RuB$_2$.  RuB$_2$ could thus be the first example of a multi-gap Type-I superconductor.      
 
\end{abstract}

\maketitle 
\section{Introduction} 
The discovery of multi-gap superconductivity in MgB$_2$ \cite{Bouquet2001, Choi2002} has led to a revival of interest and activity in the search for similar behaviour in other superconductors.  There are now several accepted candidate multi-gap superconductors such as NbSe$_2$ \cite{Boaknin2003}, $R$Ni$_2$B$_2$C ($R = $ Lu and Y) \cite{Shulga1998}, Lu$_2$Fe$_3$Si$_5$ \cite{Nakajima2008,Gordon2008}, Sr$_2$RuO$_4$ \cite{Maeno2001}, and more recently FeSe \cite{Dong2009, Amig2014}.  Multi-gap superconductors are associated with several anomalous superconducting properties.  For example, a reduced heat capacity jump at the superconducting critical temperature $\Delta C/\gamma T_c$, a non-BCS temperature dependence of the upper critical field, and a non-BCS penetration depth versus temperature.  These anomalous properties are mostly connected with Fermi surface sheets with very different characters.  This is exemplified most clearly in the case of MgB$_2$ \cite{Budko2001, Shi2014, Manzano2002, Yelland2002}.  
 
Recently OsB$_2$, which has a layered structure with puckered honeycomb Boron planes alternating with Osmium planes stacked along the $c$-axis of an orthorhombic cell, has been studied for its super-hardness as well as for its superconducting properties.  Several anomalous superconducting properties like upward curvature in the $H_c(T)$ curve, reduced heat capacity anomaly at $T_c$, non-BCS temperature dependence of the penetration depth, a small Ginzburg-Landau parameters $\kappa \sim 1$--$2$, and a first-order superconducting transition in a magnetic field have been observed for OsB$_2$ \cite{Singh2007, Singh2010}.  These properties were interpreted as signatures of two-gap superconductivity.  A fit by a two-gap model to the $T$ dependent penetration depth data gave the values $\Delta_1 \approx 1.25k_BT_c$ and $\Delta_2 \approx 1.9k_BT_c$ for the two gaps, respectively \cite{Singh2010}.  The Fermi surface of OsB$_2$ consists of a quasi-two-dimensional sheet and two nested ellipsoidal sheets \cite{Hebbache2009}.  The two gaps were argued to open on the two ellipsoidal Fermi surface sheets which are very similar in character and size \cite{Singh2010} unlike the two gaps in MgB$_2$ which open on two Fermi sheets which are very different in character \cite{Yelland2002}.  

However, an alternate view has recently been put forward for these anomalous properties of OsB$_2$ with proposal of extreme Type-I superconductivity (very small $\kappa$) and a single but highly anisotropic gap \cite{Bekaert2016}.  

RuB$_2$ is iso-structural to OsB$_2$ and is also reported to become superconducting below $T_c \approx 1.5$~K \cite{}.  Although its normal state properties have been studied in detail \cite{Singh2007}, the superconducting properties have not been explored.  Given the anomalous superconducting properties of OsB$_2$, it would be interesting to make a detailed study of the superconducting properties of RuB$_2$ to look for similar anomalous properties.  In this work we report the superconducting properties of polycrystalline samples of RuB$_2$.  We confirm that RuB$_2$ exhibits bulk superconductivity below a critical temperature $T_c = 1.5$~K\@.  The magnetization versus magnetic field data suggest Type-I superconductivity.  We estimate an electron-phonon coupling constant $\lambda_{ep} = 0.39$--$0.45$ suggesting moderate coupling superconductivity in RuB$_2$.  The extrapolated $T = 0$ critical field $H_c(0) \approx 122$--$155$~Oe is small and consistent with Type-I superconductivity.  The normalized heat capacity jump at $T_c$ was estimated to be $\Delta C/\gamma T_c \approx 1.1$, which is much smaller than the value $1.43$ expected for a single $s$-wave BCS superconductor and suggests multi-gap superconductivity.  This is confirmed by obtaining an excellent fit of the electronic specific heat data below $T_c$ to a phenomenological two-gap model.  The fit gave the gap values $\Delta_1/k_BT_c \approx 1.88$ and $\Delta_2/k_BT_c \approx 1.13$ for the two gaps.  The jump in the heat capacity at $T_c$ becomes larger in applied magnetic fields again suggesting Type-I behaviour.  These suggestions are confirmed by estimates of the Ginzburg Landau parameter $\kappa = 0.1$--$0.6$ which is smaller than the value $1/\sqrt{2} \approx 0.707$, the border between Type-I and Type-II superconductivity.  Thus RuB$_2$ could be the first multi-gap Type-I superconductor.  Additionally we calculate the band-structure and obtain the Fermi surface of RuB$_2$.  The band structure confirms metallic behaviour with majority contribution to the density of states (DOS) at the Fermi energy ($\epsilon_F$) coming from Ru $4d$ and B $2p$ orbitals.  We calculate the total DOS at $\epsilon_F = 1.17$~states/eV~f.u., where ``f.u.'' stands for ``formulae unit''.  This value is similar to the value reported for OsB$_2$.  The Fermi surface consists of $4$ sheets.  There is one quasi-two-dimensional corrugated tubular sheet and two nested ellipsoidal sheet, very similar to OsB$_2$.  An additional small $4^{\rm th}$ sheet is found which was not obtained for OsB$_2$.    \\

\begin{figure}[t]   
\includegraphics[width= 3 in]{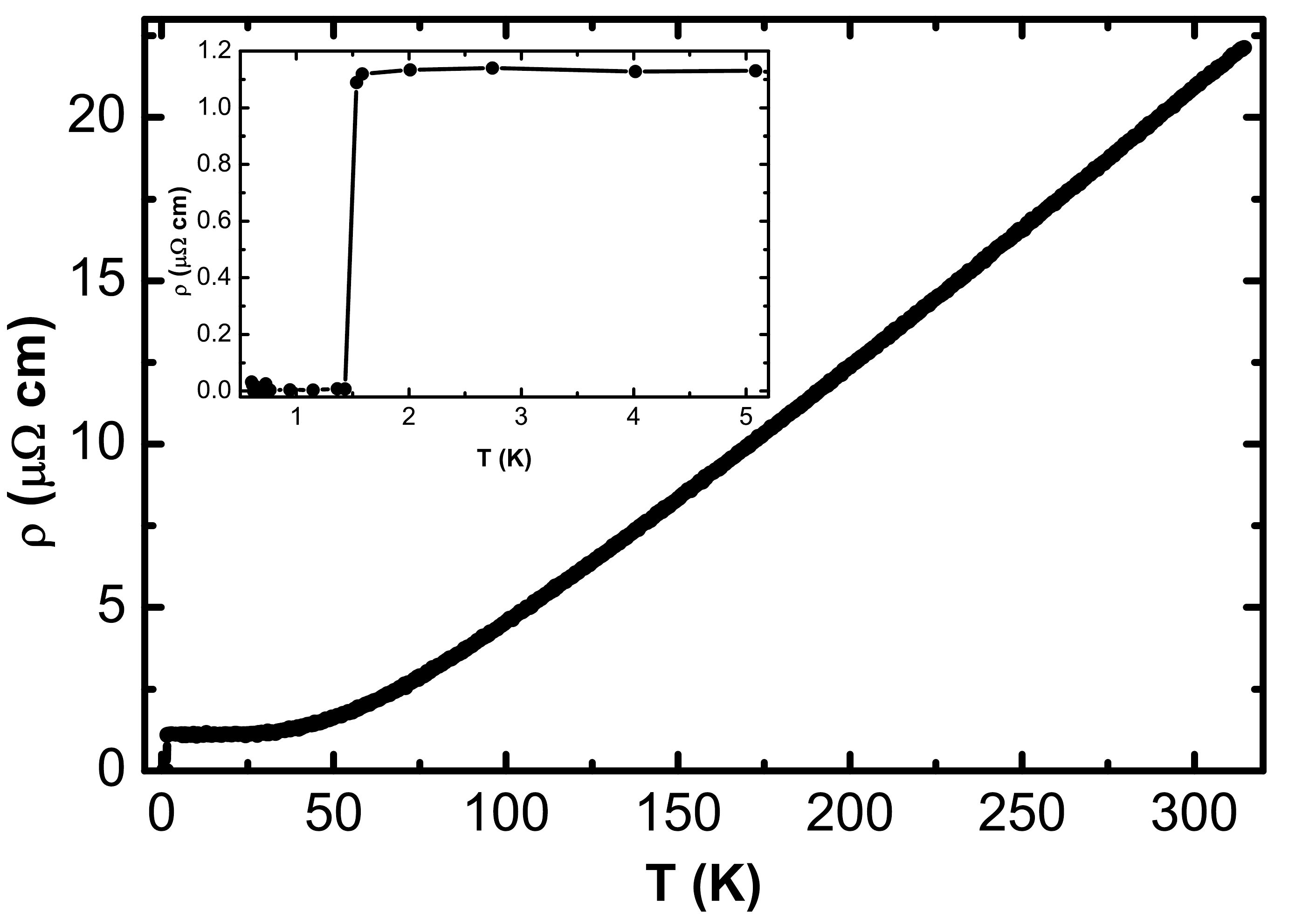}    
\caption{(Color online) The electrical resistivity $\rho$ versus temperature $T$ for RuB$_2$ measured in zero magnetic field between $T = 0.4$--$310$~K\@.  The inset shows the data below $T = 5.5$~K to highlight the abrupt drop at $T_c = 1.5$~K signalling the transition to the superconducting state.       
\label{Fig-RES}}
\end{figure}

\section{Experimental and Theoretical Methods}
Polycrystalline samples of RuB$_2$ were synthesized by arc-melting stoichiometric ratios of Ru (5N, Alfa Aesar) and B (6N, Alfa Aesar) $5$--$10$ times to promote homogeniety \cite{Singh2007}.  Powder x-ray diffraction confirmed that the synthesized material is single phase and a refinement of the powder pattern gave lattice parameters which match well with the reported values \cite{Singh2007}.  The dc magnetic susceptibility $\chi$ versus temperature data in the temperature $T$ range $T = 0.280$~K to $2$~K and magnetization $M$ versus field $H$ data at $T = 310$~mK were measured using a He3 insert in a SQUID magnetometer from Cryogenics Limited, UK.  The heat capacity $C$ data from $85$~mK to $3$~K was measured using the dilution refrigerator (DR) option of a Quantum Design Physical Property Measurement System (QD-PPMS).  The electrical transport from $300$~mK to $300$~K was measured using the He3 insert in a QD-PPMS.  The first-principles density functional theory (DFT) calculations were done using the QUANTUM-ESPRESSO code \cite{Giannozzi2009}. Electronic exchange and correlation are described using the generalized gradient approximation (GGA) using Perdew-Bruke-Ernzerhof functional \cite{Perdew1996}. 

\section{Electrical Resistivity}
The electrical resistivity $\rho$ versus temperature $T$ data for RuB$_2$ measured with an excitation current of $5$~mA in zero applied magnetic field are shown in Fig.~\ref{Fig-RES} between $T = 0.5$~K and $315$~K\@.  The $T = 315$~K value of resistivity is $\rho(315~{\rm K}) \approx 22.5~\mu\Omega$cm and the residual resistivity is $\rho(1.6~{\rm K}) \approx 1.1~\mu\Omega$cm giving a residual resistivity ratio RRR~$\approx 21$.  The inset in Fig.~\ref{Fig-RES} shows the $\rho(T)$ data below $T = 5.25$~K to highlight the sharp drop to zero resistance below $T_c = 1.5$~K signalling the onset of superconductivity in RuB$_2$. 
 
\begin{figure}[b]   
\includegraphics[width= 3.2 in]{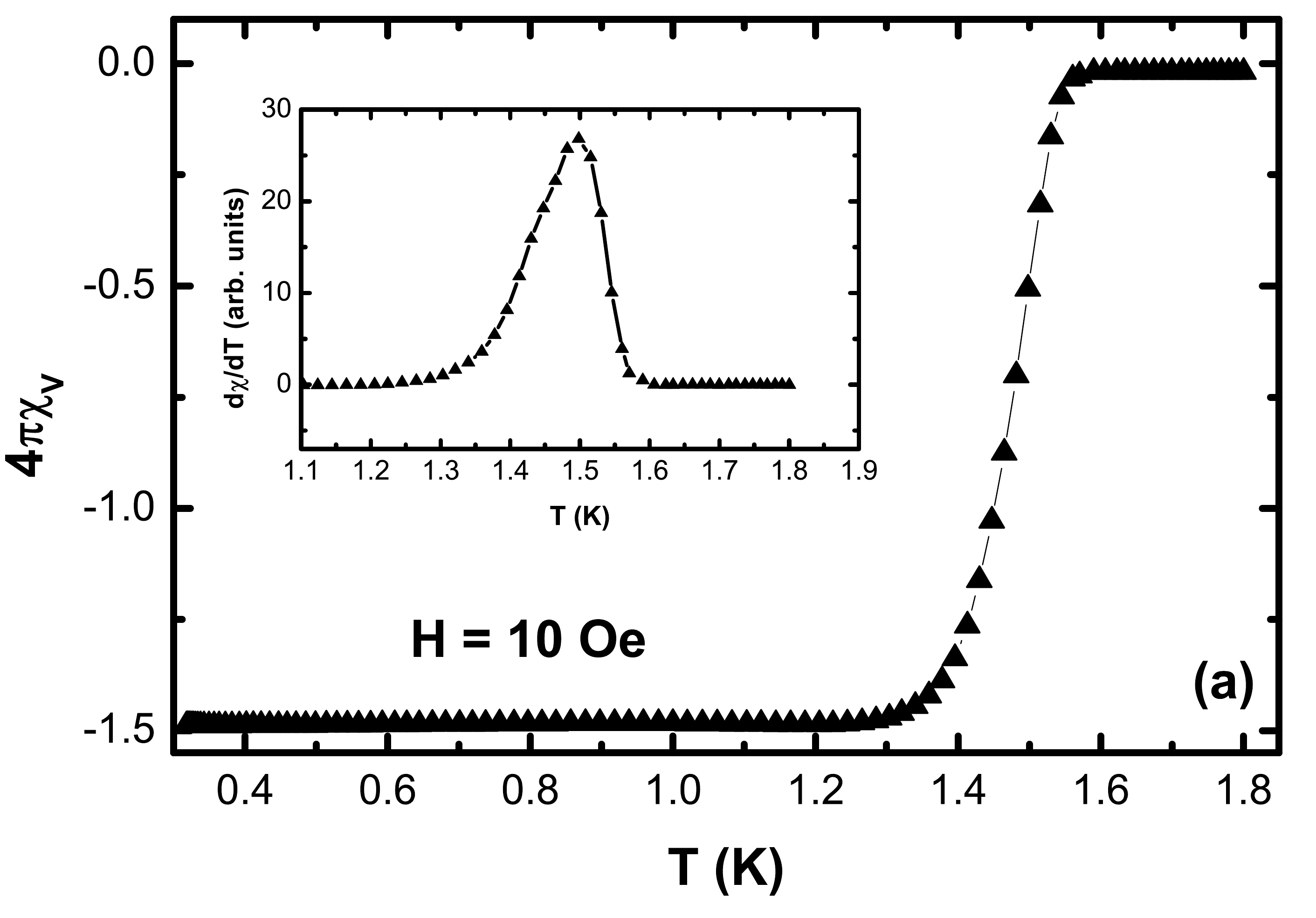}    
\includegraphics[width= 3.2 in]{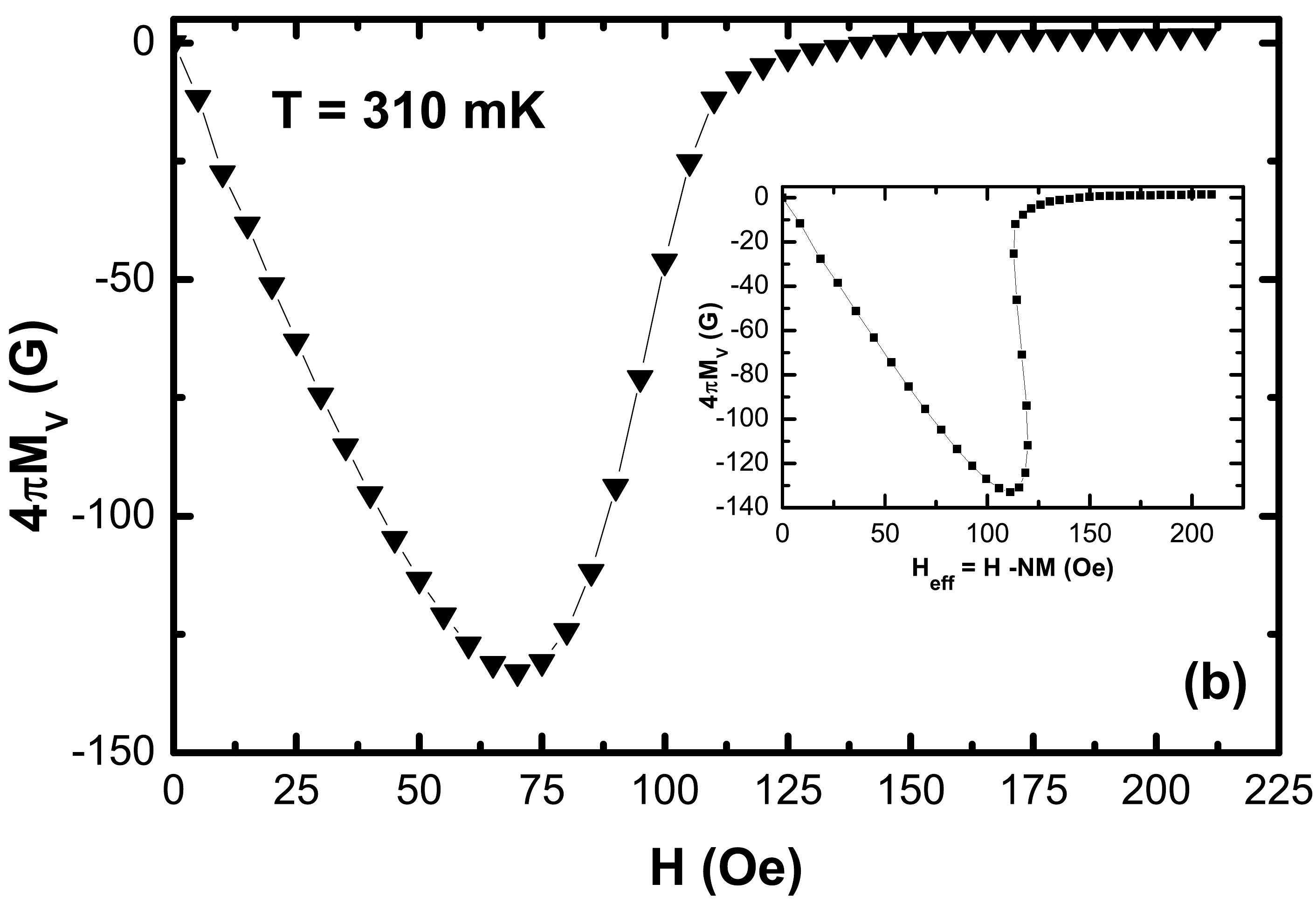}    
\caption{(Color online) (a) The temperature $T$ dependence of the zero field cooled (ZFC) dimensionless volume susceptibility $\chi_v$ in terms of the superconducting volume fraction $4\pi\chi_v$ of RuB$_2$ measured in a magnetic field $H = 10$~Oe.  At low $T$, the $4\pi\chi_v$ values are more negative than $-1$ due to demagnetization effects.  The inset shows the $d\chi/dT$ versus $T$ data to highlight the superconducting transition at $T_c = 1.5$~K\@. (b) the volume magnetization $M_v$ normalized by $1/4\pi$, versus applied magnetic field $H$ measured at $T = 310$~mK\@.  The inset shows the $4\pi M_V$ versus effective magnetic field $H_{\rm eff} = H - NM$ corrected for the demagnetization effects.   These data show behaviour typical of Type-I superconductivity.
\label{Fig-chi}}
\end{figure}

\section{Magnetic Properties}
Figure~\ref{Fig-chi} shows the results of magnetic measurements on RuB$_2$.  Fig.~\ref{Fig-chi}~(a) shows the temperature dependence of the zero-field-cooled (ZFC) volume magnetic susceptibility $\chi_v$ normalized by $1/4\pi$.  The data were measured in a field of $10$~Oe between $0.28$~K and $1.8$~K\@.  The sharp drop in $\chi_v$ to diamagnetic values below $\approx 1.55$~K confirms the onset of the superconducting state.  The inset in Fig.~\ref{Fig-chi}~(a) shows the $d\chi_v/dT$ vs $T$ data and the peak position is taken as the superconducting critical temperature $T_c = 1.5$~K\@.  The full-width-at-half-maximum (FWHM) of the peak in $d\chi_v/dT$ gives an estimate of the superconducting transition width and is $\approx 50$~mK\@.  The $\chi_v$ data have not been corrected for the
demagnetization factor $N$.  Thus, the observed value is $4\pi\chi_v = {-1\over 1-N}$ and therefore larger than $-1$ expected for $100\%$ superconducting volume fraction.  Assuming $100\%$ superconducting volume fraction we estimate $N \approx 0.32$ from the data shown in Fig.~\ref{Fig-chi}~(a).  However, often in polycrystalline samples the superconducting fraction is smaller than $100\%$ and to estimate the actual superconducting fraction one needs the value of $N$.  For idealized shapes of measured samples, $N$ has been calculated.  For example, $N = 1/3$ for a sphere and $1$ for a ellipsoid of revolution.  Our sample is an irregular shaped piece which looks like a squashed ellipsoid with dimensions $a \approx b = 1.61$~mm $\neq c = 1.35$~mm, broken from an arc-melted button.  We therefore approximate our irregular shaped sample with a prolate ellipsoid with $c/a \approx 0.83$.  For such an object, $N \approx 0.38$~~ \cite{Poole}.   Using this value of $N$ we find a superconducting volume fraction of $\approx 90\%$.

Figure~\ref{Fig-chi}~(b) shows the volume magnetization $M_v$ normalized by $1/4\pi$ versus magnetic field $H$ for RuB$_2$ measured at a temperature $T =3 10$~mK, well inside the superconducting state.  The shape of the $4\pi M_v$ vs $H$ data are very different from those expected for typical Type-II superconductors but are similar to that expected for a Type-I superconductor with demagnetization factors.  To account for demagnetization effects the magnetization can be plotted versus an effective magnetic field $H_{\rm eff} = H-NM$.  This has been done using the $N \approx 0.39$ estimated above and the resulting $M(H_{\rm eff})$ data are shown in Figure~\ref{Fig-chi}~(b)~inset.  These data look like the behaviour expected for a Type-I superconductor.   The slight negative slope of the data at the transition most likely occurs from a slightly overestimated $N$.     
  
\section{Heat Capacity}  

\begin{figure}[t]   
\includegraphics[width= 3 in]{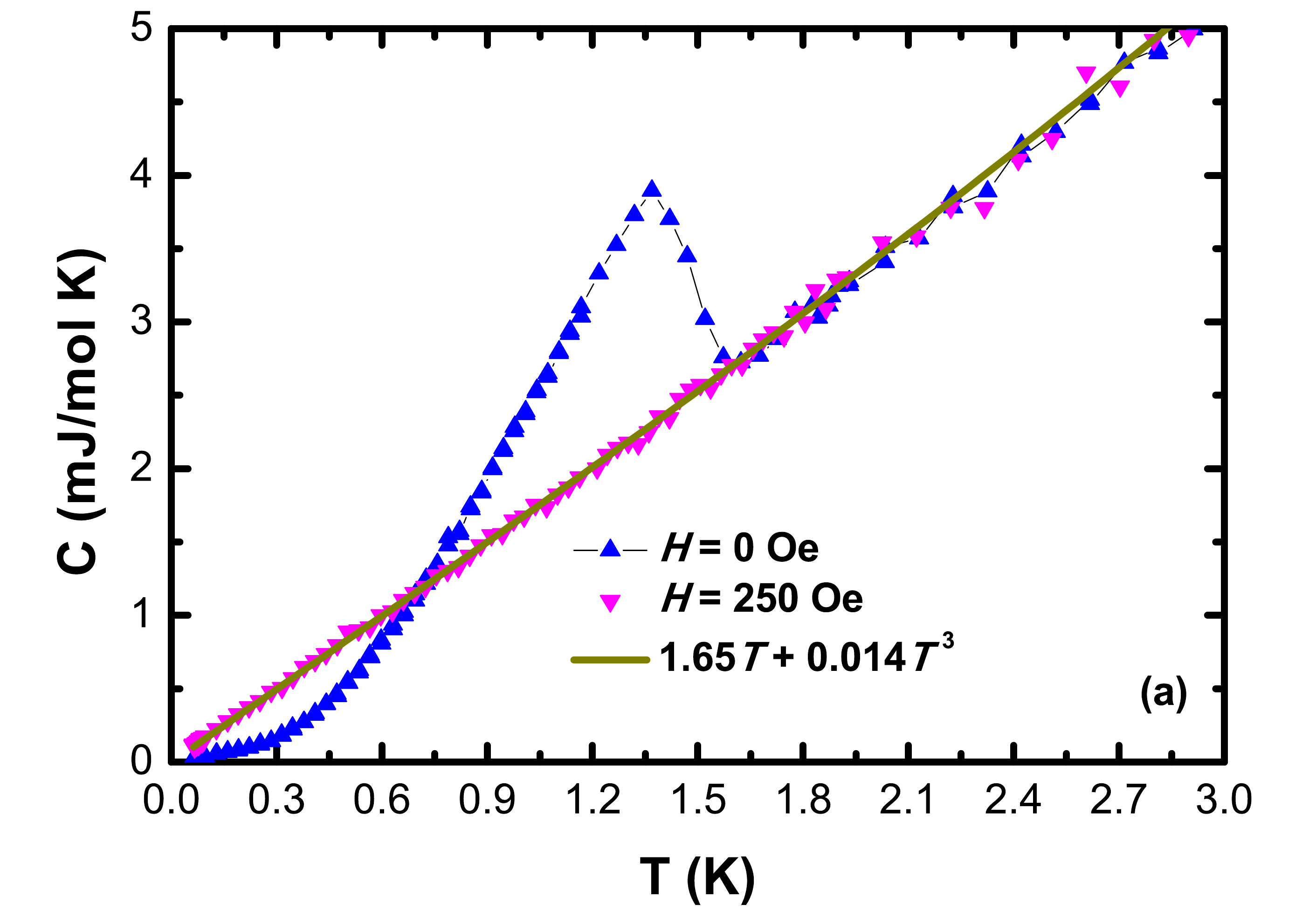}    
\includegraphics[width= 3 in]{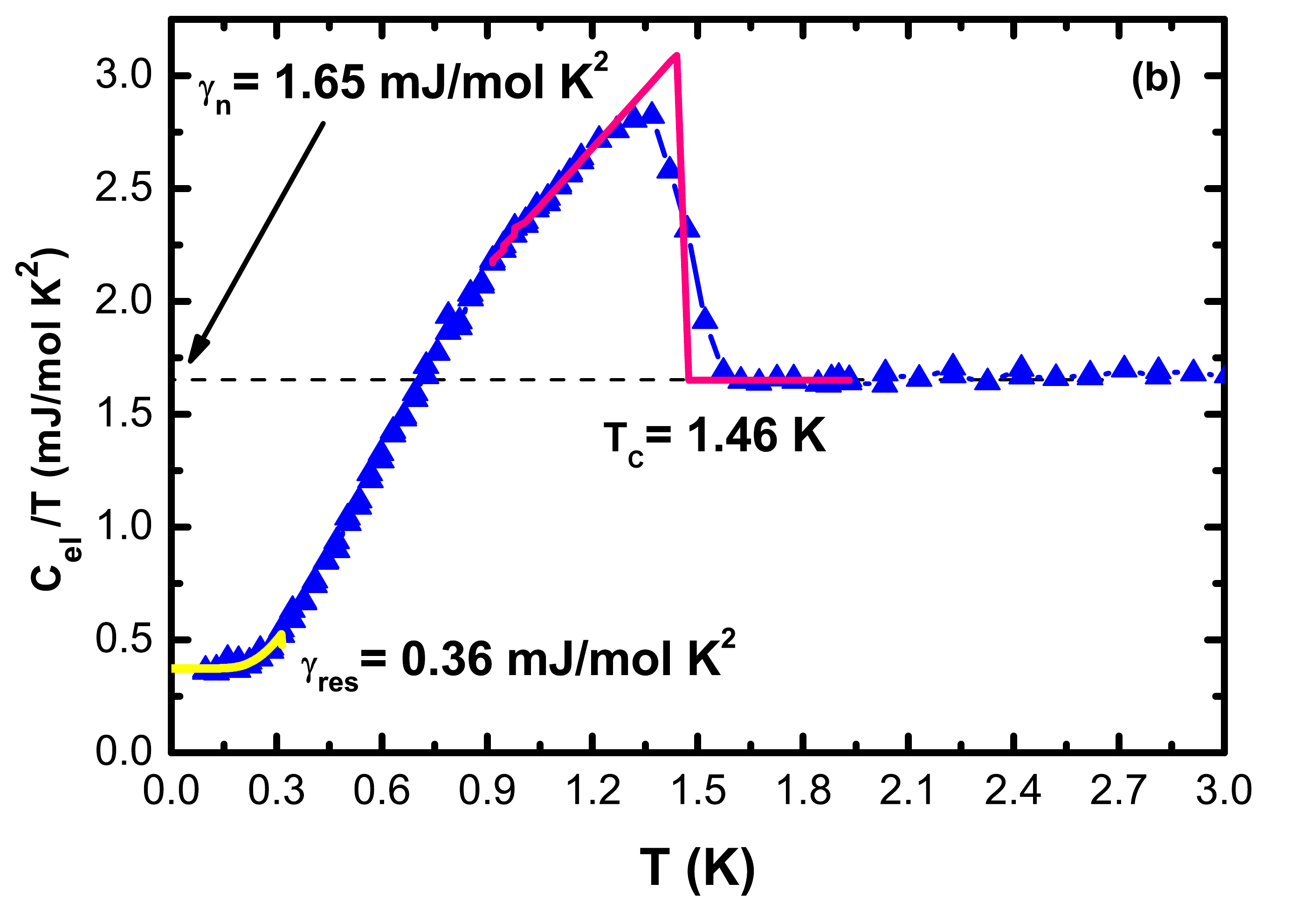} 
\includegraphics[width= 3 in]{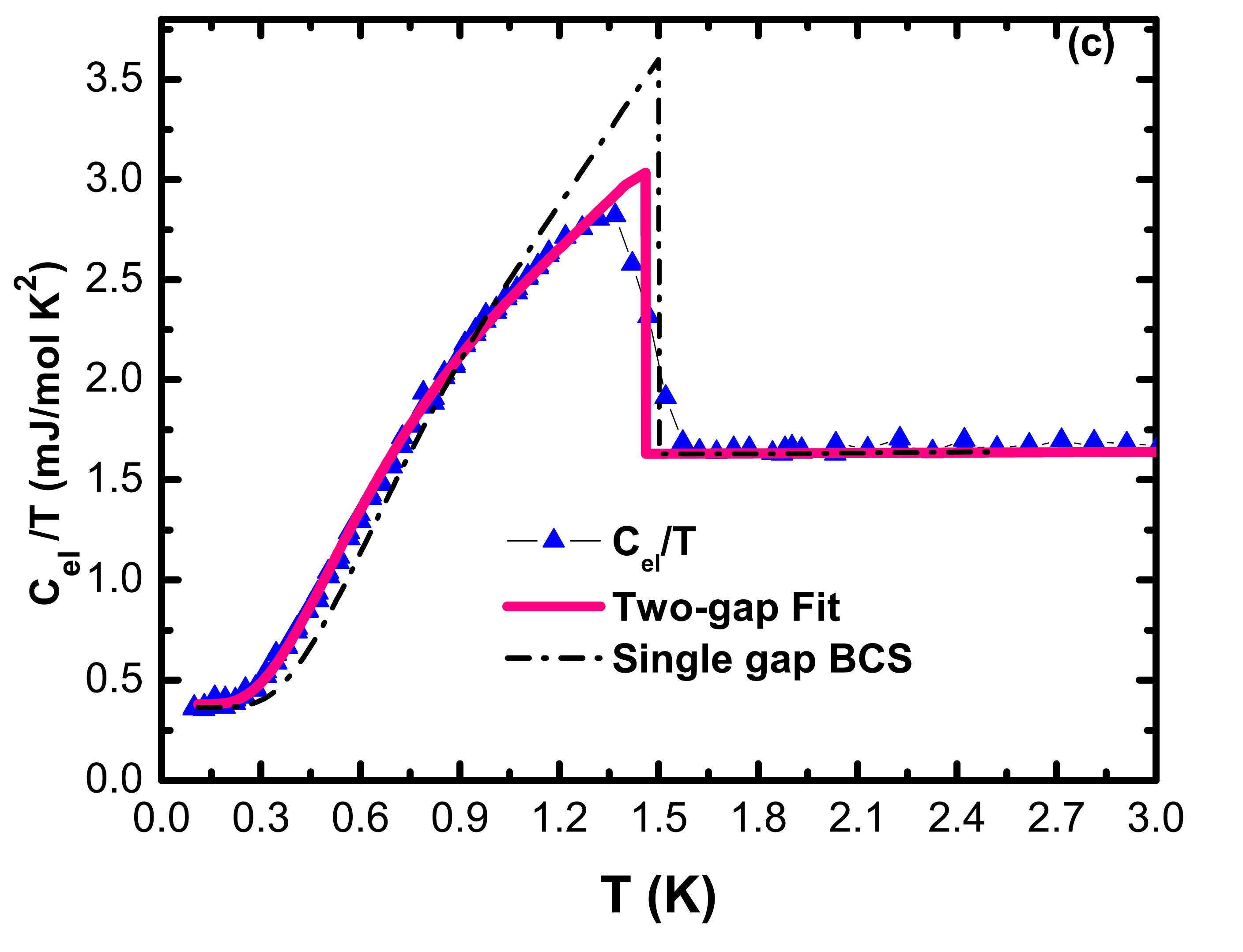}    
\caption{(Color online) (a) Specific heat $C$ versus $T$ for RuB$_2$ measured in magnetic fields $H= 0,$ and $250$~Oe.  (b) The electronic specific heat divided by temperature $C_{el}/T$ versus $T$ for RuB$_2$.  An equal entropy construction is shown to give a $T_c = 1.46$~K and $\Delta C/\gamma T_c = 1.1$, where $\gamma = \gamma_n - \gamma_{res}$.  (c) A two-gap model fit (solid curve) to the $C_{el}$ data and expectation for a single BCS gap with $T_c = 1.5$~K\@ (see text for details).         
\label{Fig-CP-new}}
\end{figure}

Figure~\ref{Fig-CP-new}~(a) shows the specific heat $C$ versus $T$ data for RuB$_2$ measured between $T = 85$~mK and $3$~K in magnetic fields $H = 0$~Oe and $H = 250$~Oe.  A sharp anomaly near $T_c = 1.5$~K in the $H = 0$ data confirms bulk superconductivity in RuB$_2$.  The data at $H = 250$~Oe doesn't show any signature of superconductivity and will be used as the normal state data.  We will later show that this field is indeed much higher than the estimated critical field.  The $C(T)$ data at $H = 250$~Oe were fit by the expression $C = \gamma_n T + \beta T^3$ where $\gamma_n$ is the normal state Sommerfeld coefficient and the second term is the contribution from the lattice.  The fit shown as the solid curve through the $H = 250$~Oe data in Fig.~\ref{Fig-CP-new}~(a) gave the values $\gamma_n = 1.65(2)$~mJ/mol~K$^2$ and $\beta = 0.014(2)$~mJ/mol~K$^4$.  This value of $\beta$ gives a Debye temperature of $\theta_D = 720(30)$~K which is similar to the value found previously \cite{Singh2010}.  The lattice contribution $\beta T^3$ to the total specific heat $C(T)$ can be subtracted to get the electronic contribution $C_{el}(T)$.  The $C_{el}(T)$ so obtained is shown in Fig.~\ref{Fig-CP-new}~(b).  The sharp anomaly at $T_c$ as well as the exponential fall at the lowest temperatures expected for $s$-wave superconductors is clearly visible.  We also note that $C_{el}$ tends to a finite value as $T \rightarrow 0$ suggesting some non-superconducting fraction in the sample.  The data below $\approx 0.3$~K were fit by the expression $ C_{el}/T = \gamma_{res} + (A/T)exp(-\Delta/T)$, where $\gamma_{res}$ is the residual Sommerfeld coefficient from the non-superconducting fraction of the sample and the second term is a phenomenological exponential decay expected for a gapped system.  The fit shown as the solid curve through the data below $T \approx 0.3$~K in Fig.~\ref{Fig-CP-new}~(b) gives the value $\gamma_{res} = 0.36$~mJ/mol~K$^2$.  With the total $\gamma_n = 1.65$~mJ/mol~K$^2$ and the residual non-superconducting $\gamma_{res} = 0.36$~mJ/mol~K$^2$, the superconducting contribution becomes $\gamma_s = 1.29$~mJ/mol~K$^2$.  This suggests that $\approx 22\%$ of the sample volume is non-superconducting.  

We can now analyze the specific heat jump height at $T_c$.  The jump $\Delta C$ at $T_c$ is normalized as $\Delta C /\gamma T_c$, where $\gamma$ is the Sommerfeld coefficient of the superconducting part.  The superconducting transition can be broadened and the jump height suppressed in real materials due to a distribution of $T_c$ arising from sample inhomogeneities or disorder.  To get a better estimate of $\Delta C$ and $T_c$ we use an entropy-conserving construction.  In such a construction the $C_{el}$ data just below the maximum of the anomaly is fit by a polynomial and extrapolated to higher temperatures.  The entropy is then evaluated and equated to the normal state entropy $\gamma_n T_c$.  Such a construction gave the jump height $\Delta C/T_c = 3.07 - 1.65 = 1.42$~mJ/mol~K$^2$ and $T_c = 1.46$~K as shown in the Fig.~\ref{Fig-CP-new}~(b).  The $T_c$ found by this entropy-conserving construction is quite close to the onset temperature $1.5$~K indicating the sharp transition and suggesting a very good sample quality with very little disorder and inhomogeneities.  Using the above $\Delta C/T_c = 1.42$~mJ/mol~K$^2$ and the superconducting contribution $\gamma_s = 1.29$~mJ/mol~K$^2$ we estimate $\Delta C/\gamma_s T_c = 1.44/1.29 \approx 1.12$.  This value is much smaller than the value $1.43$ expected for a single-gap $s$-wave superconductor.  The reduced value of $\Delta C /\gamma T_c $ is similar to observations for MgB$_2$ \cite{Budko2001} and OsB$_2$ \cite{Singh2007, Singh2010} and suggests multi-gap superconductivity.  

To confirm this possibility we have attempted to fit our $C_{el}(T)$ data below $T_c$ to a phenomenological two-gap model as has been reported for example for MgB$_2$ \cite{Bouquet2001}.  The $T = 0$ value of the two superconducting gaps $\Delta_1$ and $\Delta_2$, the critical temperature $T_c$, and the fractional contribution of the first band $x$ were the three fit parameters.  An excellent fit, shown in Fig.~\ref{Fig-CP-new}~(c) as the solid curve through the $C_{el}/T$ data below $T_c$, was obtained with the fit parameters $\Delta_1/k_BT_c \approx 1.88$, $\Delta_2/k_BT_c \approx 1.13$, $T_c \approx 1.47$~K, and $x = 0.58$.   If we compare the values of the two gaps we estimate above to the single band BCS value $\Delta/k_BT_c = 1.76$ we see that our values agree with the theorem that for a two-gap superconductor one of the gaps will always be larger than the BCS value while the second gap will always be smaller \cite{Kresin1990}.  For comparison, we also show in Fig.~\ref{Fig-CP-new}~(c) the simulated data for superconductor with a single BCS gap with $T_c = 1.5$~K which clearly doesn't match the data. Thus, the heat capacity data in Fig.~\ref{Fig-CP-new} strongly indicate that RuB$_2$ could be a two-gap superconductor.

\begin{figure}[t]   
\includegraphics[width= 3 in]{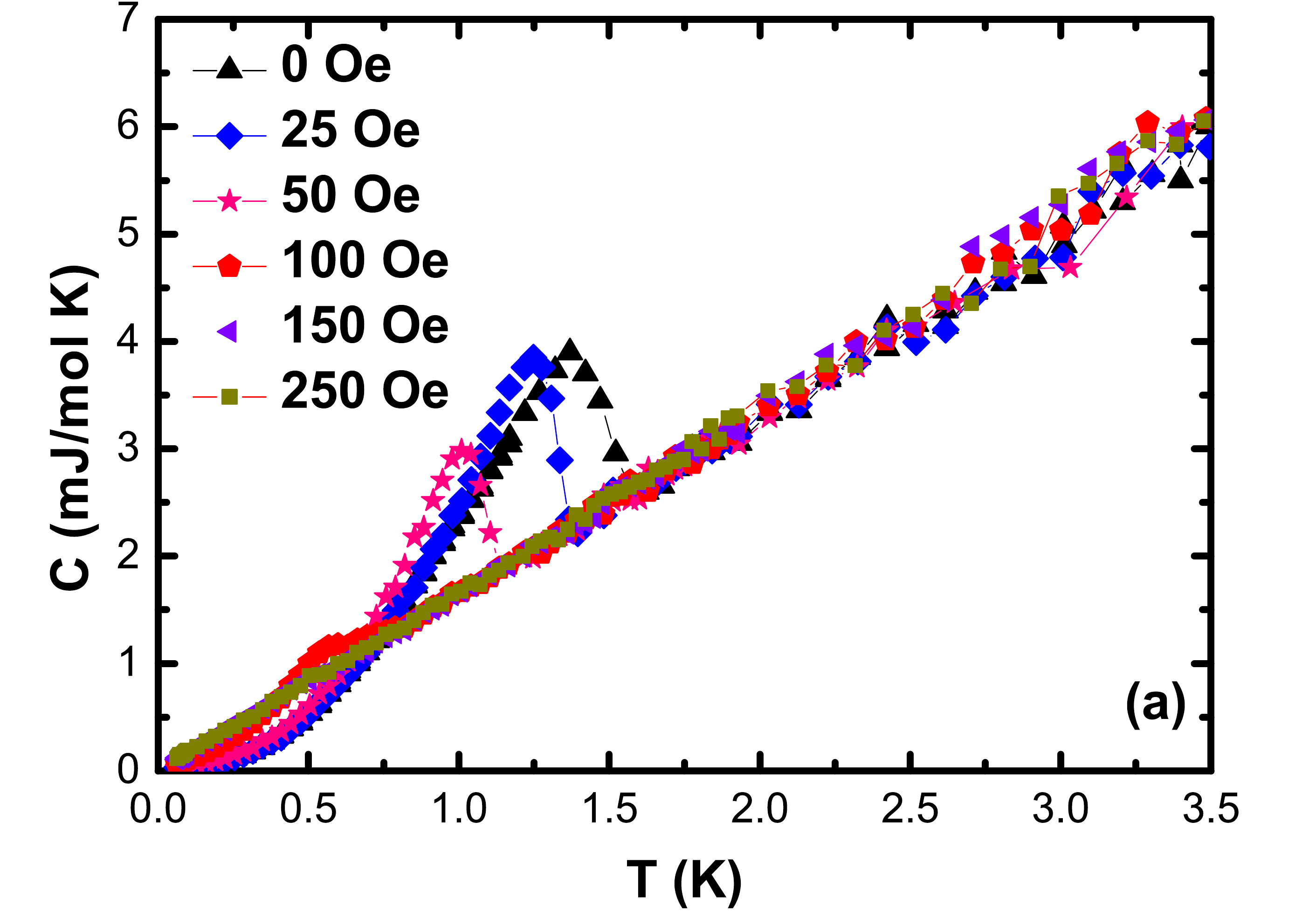}    
\includegraphics[width= 3 in]{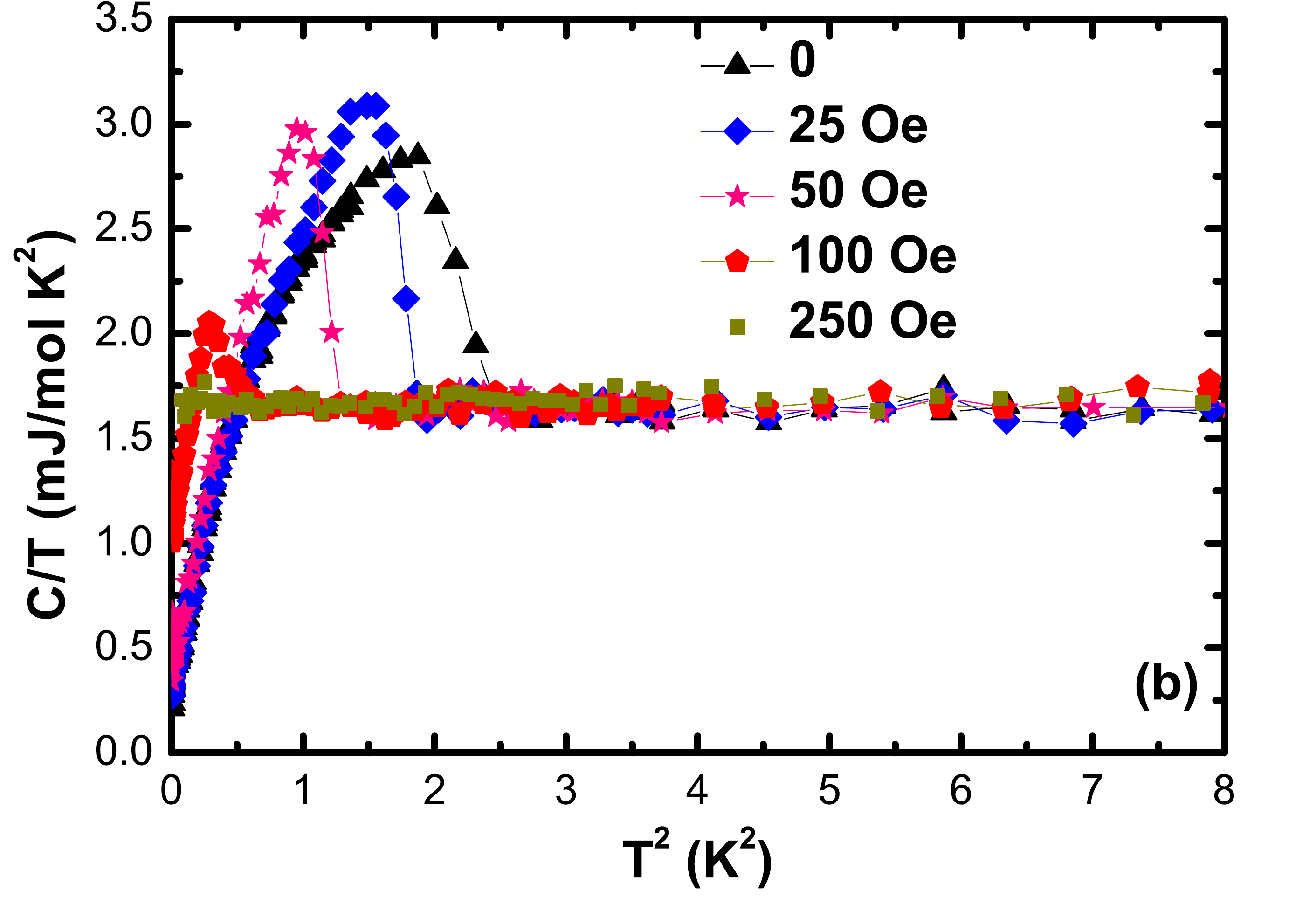}    
\caption{(Color online) (a) Specific heat $C$ versus $T$ for RuB$_2$ measured in various magnetic fields $H$.  (b) $C$ divided by temperature $C/T$ versus $T^2$ for RuB$_2$ at various $H$.  The solid curve through the data is a fit by the expression $C = \gamma T + \beta T^3$.  The peak height at $T_c$ in zero field is characterised by the ratio $\Delta C/\gamma T_c$ and is estimated to be $0.8$ for RuB$_2$.        
\label{Fig-CP}}
\end{figure}

Figure~\ref{Fig-CP}~(a) shows the specific heat $C$ versus $T$ data for RuB$_2$ measured between $T = 85$~mK and $3.5$~K at various applied magnetic fields $H$.  All data were measured by cooling in zero field to the lowest temperature and then measuring while warming up in the desired magnetic field.  As expected, the SC transition is pushed to lower temperatures in increasing fields and is not observed down to the lowest temperature for fields $H \geq 250$~Oe.  The specific heat divided by temperature $C/T$ versus $T$ at various magnetic fields is plotted in Fig.~\ref{Fig-CP}~(b).  From Fig.~\ref{Fig-CP}~(b) we observe that the magnitude of the peak at $T_c$ initially increases in a magnetic field.  In a magnetic field the transition for a Type-I superconductor becomes first-order.  Thus, one should in principle observe a diverging anomaly at $T_c$.  In real materials however, the anomaly is broadened due to sample inhomogeneity and as a consequence the anomaly looks like a jump larger than that in zero field.  Thus the observed behaviour in Fig.~\ref{Fig-CP}~(b) also points to Type-I superconductivity in RuB$_2$.  This is similar to what was observed for OsB$_2$\cite{Singh2010} and for other Type-I superconductors like ScGa$_3$ and LaGa$_3$ \cite{Svanidze2012} and YbSb$_2$ \cite{Zhao2012}.  This is consistent with the magnetization data of Fig.~\ref{Fig-chi}~(b) inset which also suggest Type-I superconductivity.

The above value of $\gamma$ can be used to estimate the density of states at the Fermi energy ($\epsilon_F$) for both spin directions $N(\epsilon_F)$ by using the expression $\gamma = {\pi^2\over 6}k_B^2N(\epsilon_F)$.  Using $\gamma = 1.65$~mJ/mol~K$^2$ we obtain $N(\epsilon_F) \approx 1.40$~states/eV~f.u.  We will compare this value with estimations from band structure calculations later.

\section{Superconducting Parameters} 
\begin{figure}[t]   
\includegraphics[width= 3 in]{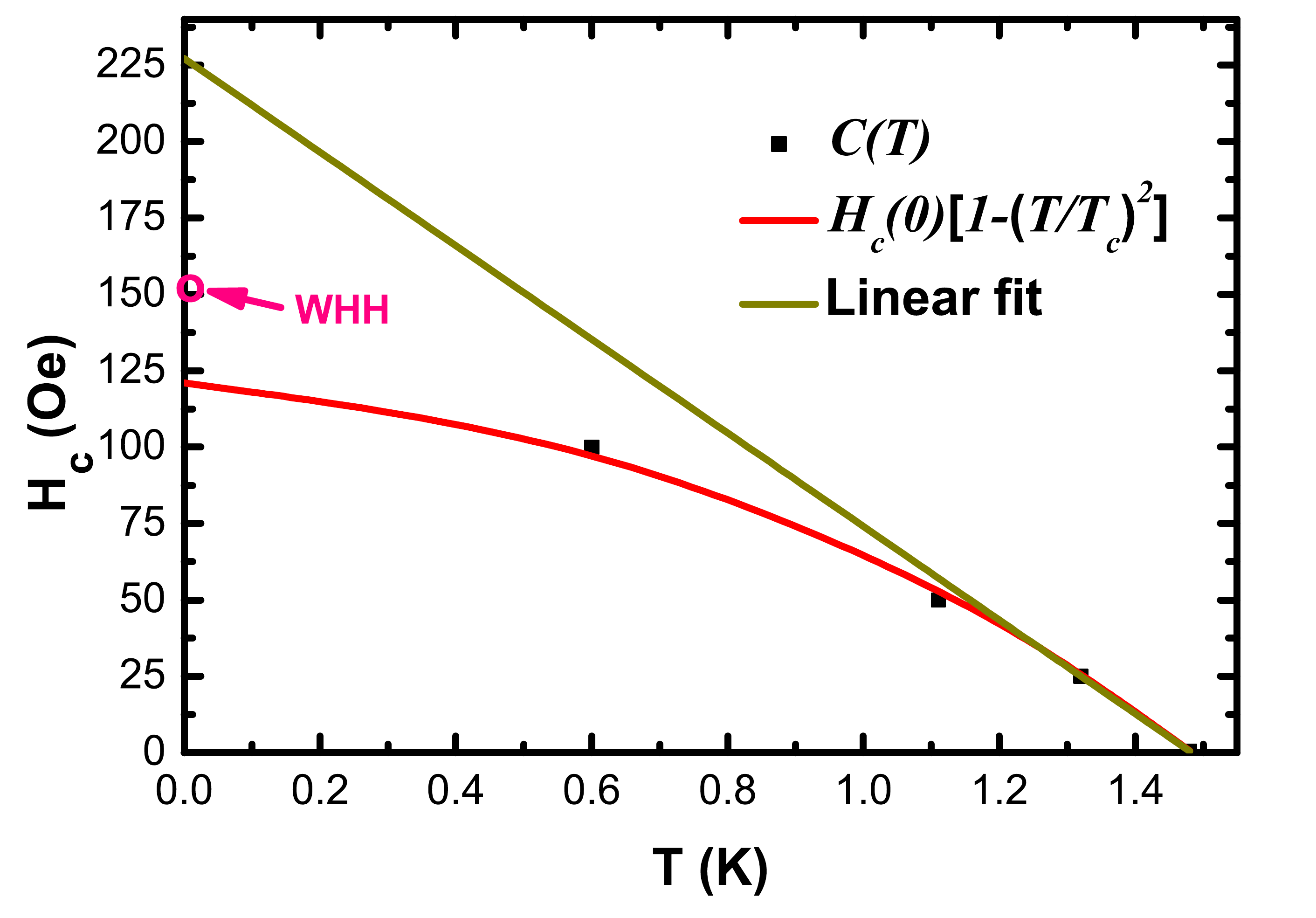}    
\caption{(Color online) The critical field $H_C$ versus $T$ data extracted from the heat capacity $C$ versus temperature $T$ at various $H$.  The solid curve is a fit to the phenomenological BCS expression (see text).  The linear curve is a linear fit to the data close to $T_c$ and extrapolated to $T = 0$.  The WHH value of the $H_c(0) = 0.69 \times$ this extrapolated value.     
\label{Fig-HC}}
\end{figure}

The $C(T,H)$ data presented above were used to extract the critical temperature at various magnetic fields.  The critical field $H_c$ versus $T$ data thus obtained is shown in Fig.~\ref{Fig-HC}.  The data were fit by the phenomenological expression $H_c(T) = H_c(0)[1-({T\over T_c})^2]$ with $H_c(0)$ and $T_c$ as fitting parameters, where $H_c(0)$ is the zero temperature critical field.  The fit, shown as the solid curve through the data in Fig.~\ref{Fig-HC}, extrapolated to $T = 0$ gave the values $H_c(0) = 122$~Oe and $T_c = 1.48$~K\@.  The excellent fit to the above expression suggests BCS superconductivity in RuB$_2$.  We note that the above expression is strictly valid close to $T = 0$ where the data density is very less.  Another estimate of $H_c(0)$ can be made using the Werthamer-Helfand-Hohenberg (WHH) formula which gives $H_c(0) = -0.693T_c({dH_c \over dT}|_{Tc})$.  The linear fit extrapolated to $T = 0$ is shown in Fig.~\ref{Fig-HC}.  The WHH estimate for $H_c(0)$ can be obtained by multiplying the extrapolated value with $0.693$.  This gives $H_c(0) \approx 153$~Oe and is shown in Fig.~\ref{Fig-HC}.    

The electron-phonon coupling $\lambda_{\rm ep}$ can be estimated using McMillan's formula \cite{McMillan}, which relates the superconducting transition temperature $T_c$ to $\lambda_{\rm ep}$, the Debye temperature $\theta_D$, and the Coulumb pseudopotential $\mu^*$.  This formula can be inverted to get $\lambda_{\rm ep}$ in terms of the other parameters, 

$$ \lambda_{\rm ep}={1.04+\mu^*{\rm ln}({\theta_D \over 1.45T_c})\over (1-0.62\mu^*){\rm ln}({\theta_D \over 1.45T_c})-1.04}~~.$$ 

Using, $\theta_D = 700$~K obtained from heat capacity measurements above and using $T_c = 1.5$~K, we get $\lambda_{\rm ep} = 0.37$ and $0.45$ for $\mu^* = 0.1$ and $0.15$, respectively.  These values are slightly smaller than values obtained for OsB$_2$ \cite{Singh2007} consistent with a slightly smaller $T_c$ compared to OsB$_2$.  These values of $\lambda_{\rm ep}$ suggest moderate-coupling superconductivity in RuB$_2$.  The corresponding value for MgB$_2$ is $\lambda_{\rm ep} \approx 1$~~ \cite{Pickett}.  

We now estimate the $T = 0$ values of the penetration depth $\lambda(0)$ and coherence length $\xi(0)$.  RuB$_2$ has 2 formulae units per unit cell.  This means that there are $4$ electrons in one unit cell volume $V = 53.84$~\AA$^3$.  Therefore, the electron density is $n = 4/V = 7.4 \times 10^{-2}$~\AA$^{-3}$.  Assuming a spherical Fermi surface, we can use the above value of $n$ to estimate the Fermi wave-vector $k_F = (3n\pi^2)^{1/3} = 1.3$~\AA$^{-1}$.  The London penetration depth is given by $\lambda(0) = (m^*/\mu_0ne^2)^{1/2}$, where we take the effective mass $m^*$ as the free electron mass $m_e$.  Putting in values gives us $\lambda(0) \approx 47$~nm.  The BCS coherence length can be estimated using the expression $\xi = {0.18\hbar^2 k_F \over k_BT_cm^*} \approx 0.45~\mu$m.  The Ginzburg Landau (GL) parameter can now be estimated as $\kappa = \lambda(0)/\xi \approx 0.1$ which is much smaller than the value $1/\sqrt{2} \approx 0.707$ separating Type-I and Type-II superconductivity.  The above value of $\kappa$ suggests that RuB$_2$ is an extreme Type-I superconductor.  This is consistent with the low $H_c$ and the $M(H_{\rm eff})$ data presented above.  The mean free path $l$ can be estimated using the expression $l = v_F\tau$, where the Fermi velocity is $v_F = \hbar k_F/m^*$ and the scattering time is given by the expression for the Drude conductivity $\tau = m^*/ne^2\rho$.  Using $m^* = m_e$ and the residual resistivity value $\rho(1.6~{\rm K}) = 1.1~\mu\Omega$~cm, we estimate $l \approx 72$~nm.  From the above estimates of $\xi$ and $l$ we conclude that $\xi >> l$, making RuB$_2$ a dirty limit superconductor.  For a dirty limit superconductor we can make another estimate of the GL parameter as $\kappa = 0.75\lambda(0)/l \approx 0.66 < 0.707$, again consistent with Type-I behaviour.

\section{Band Structure and Fermi Surface}  
RuB$_2$ crystallises in the orthorhombic crystal system, space group $Pmmn$ (no. 59).  Each unit cell contains two formula units (two Ru atoms and four B atoms).  The ionic and lattice relaxation were performed to optimize the crystal structure by using variable cell relaxation.  We have used an energy cutoff of $55$~Ry for the plane wave basis.  The Brillouin zone integration is conducted with a $11\times18\times13$ Monkhorst-pack grid for the K-point sampling. In the optimized crystal structure, the forces on all the atoms are less than $10^{-4}$~Ry/au. The calculated lattice parameters of optimized RuB$_2$ compound along with the experimental values are tabulated in Table~\ref{Table-Lattice Parameters}.  The calculated lattice parameters are within $1\%$ of the experimental values \cite{Singh2007}.

\begin{table}
\caption{ Lattice parameters obtained from relaxing the experimental unit cell of RuB$_2$}
\begin{ruledtabular}
\begin{tabular}{|c|c|c|c|}
Lattice Parameters(\AA) & Experimental & Calculated& $\%$Error   \\ \hline  
$a$ & $4.644795$ & $4.66487$ & $0.43$ \\
$b$ & $2.865153$ & $2.89674$ & $1.1$  \\
$c$ & $4.045606$ & $4.05224$  & $0.16$  \\ 
\end{tabular}
\end{ruledtabular}
\label{Table-Lattice Parameters}
\end{table} 

\begin{figure}[t]   
\includegraphics[width= 3 in]{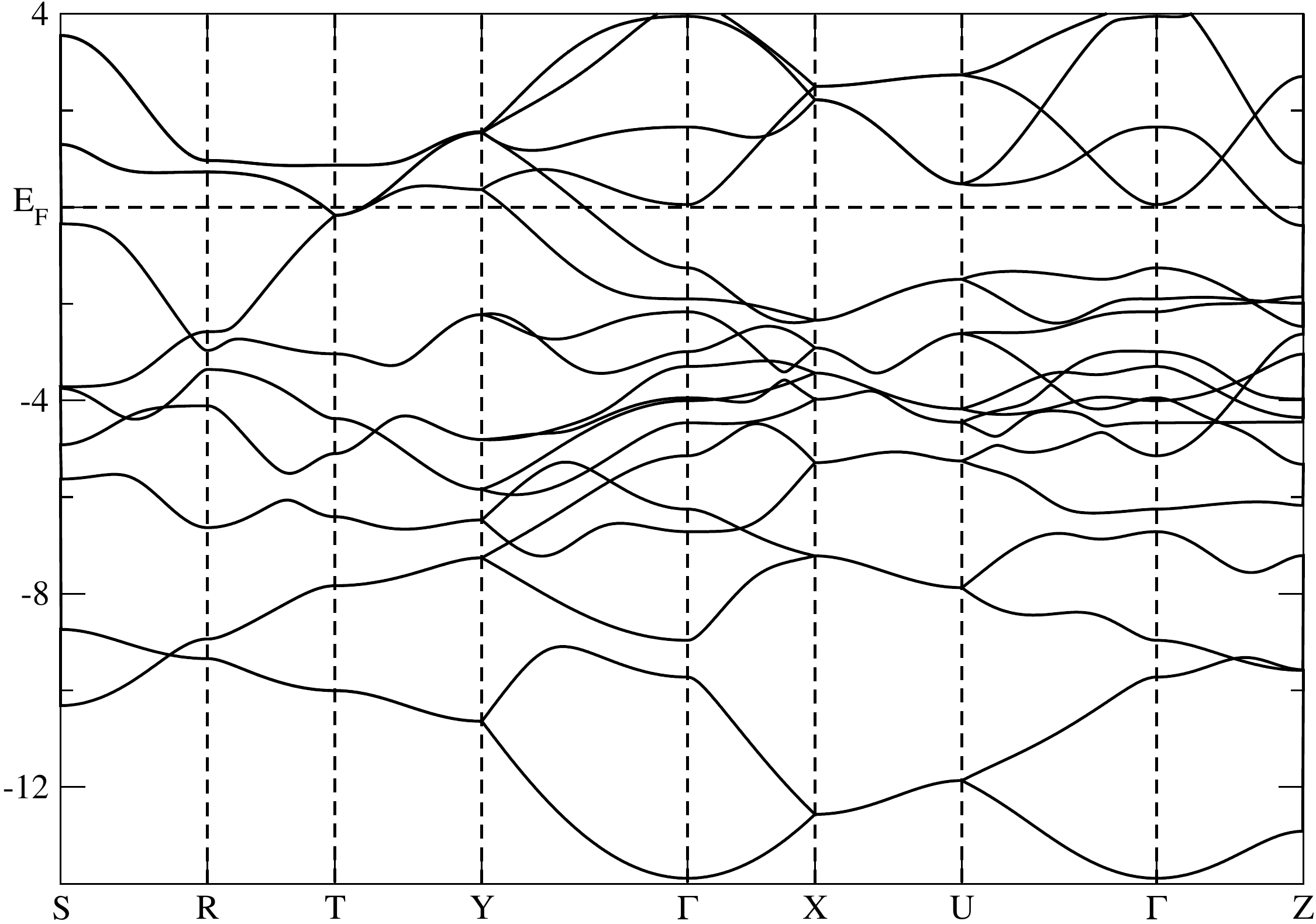}    
\caption{(Color online) The calculated electronic band structure of orthorhombic RuB$_2$ along high symmetric points. $E_F$ represents the Fermi level, which is set at $0$~eV\@.      
\label{Fig-BS}}
\end{figure}

\begin{figure}[t]   
\includegraphics[width= 3 in]{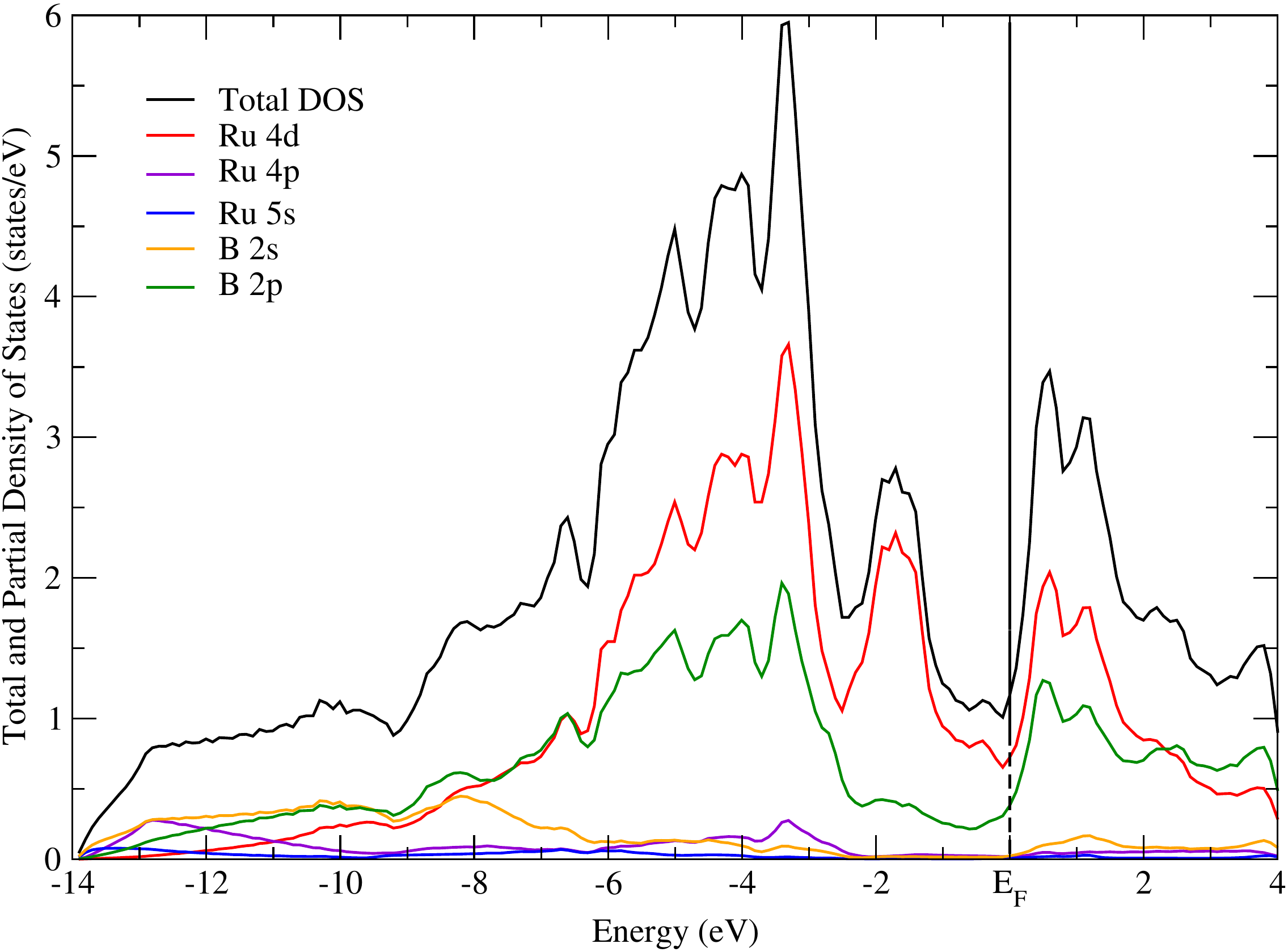}    
\caption{(Color online) Calculated total density of states (DOS) and partial density of states (PDOS) for RuB$_2$.  $E_F$, represents the Fermi energy and is set at $0$~eV\@.      
\label{Fig-DOS}}
\end{figure}

\begin{figure}[t]   
\includegraphics[width= 2. in]{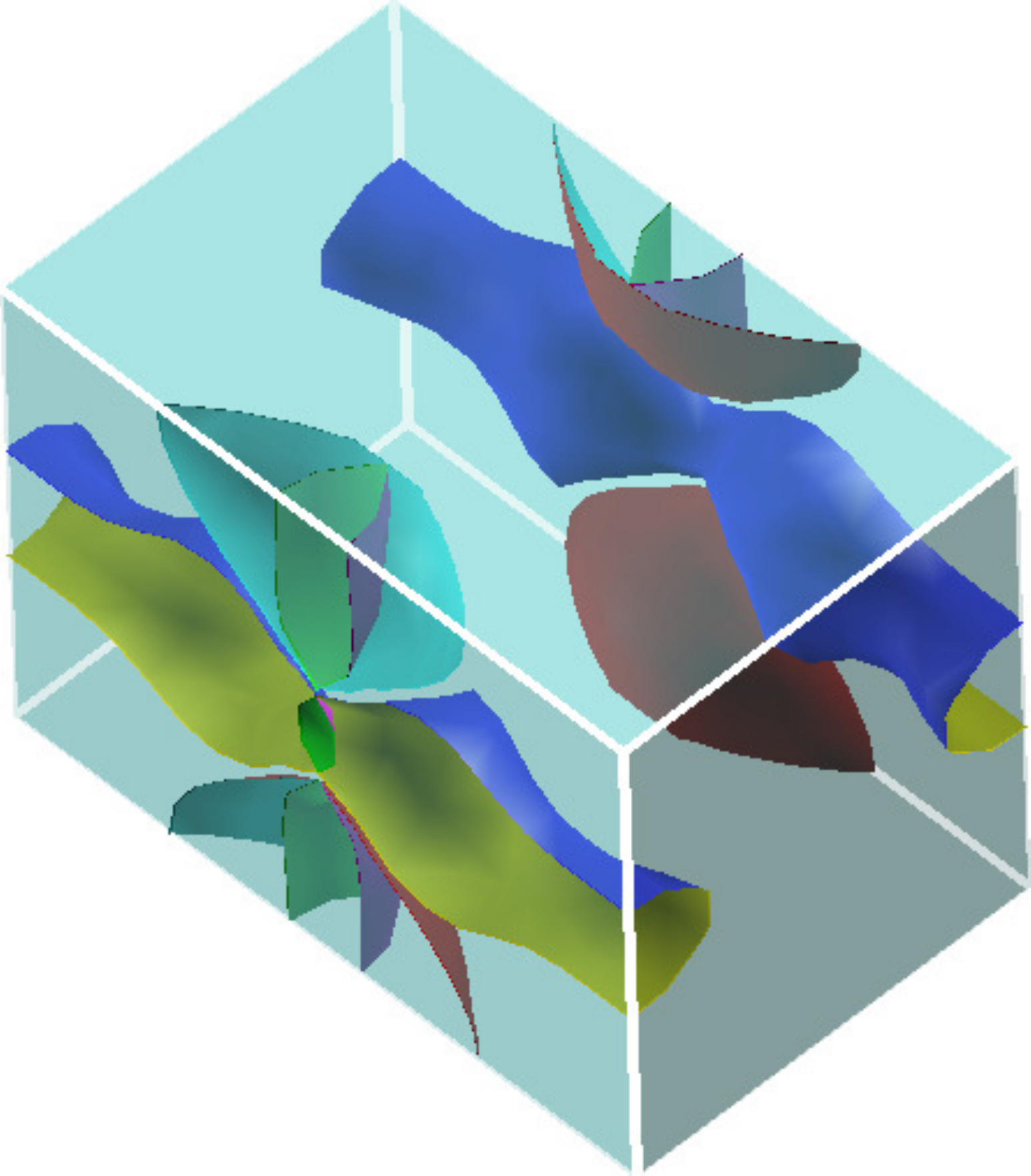}    
\caption{(Color online) The merged Fermi surface (FS) for RuB$_2$ consisting of $4$ different sheets.  The parallelepiped is in the first Brillouin zone.     
\label{Fig-FS}}
\end{figure}

The electronic band structure of RuB$_2$ is shown in Fig.~\ref{Fig-BS}.  It can be seen that several energy bands are crossing the Fermi level $E_F$ confirming that RuB$_2$ is a metal.  Figure~\ref{Fig-DOS} shows the total and partial density of states (DOS) in units of states/eV showing the contribution of individual elements and orbitals to the DOS at various energies measured from the Fermi energy $E_F$.  From Fig.~\ref{Fig-DOS} it can be seen that the $5d$-orbital of Ru and the $2p$-orbital of B make the main contributions to the density of states in the vicinity of the Fermi level.  The total DOS at $\epsilon_F$ is found to be $ N(\epsilon_F) = 1.15$~states/eV~f.u. for both spin directions.  This value is slightly smaller than the value $ N(\epsilon_F) = 1.40$~states/eV~f.u. estimated from experimental value of $\gamma$.  An estimate for the electron-phonon coupling constant $\lambda_{ep}$  can be made using the following relation:\\
 $ N(\epsilon_F)$ from heat capacity $= ( N(\epsilon_F)$ from band structure )$(1+\lambda_{ep})$.\\
   A comparison of the above experimental and theoretical values of $ N(\epsilon_F)$ gives $\lambda_{ep} \approx 0.22$ which is close but slightly smaller than the values obtained above using McMillan's formula.  

We have also obtained the Fermi surface for RuB$_2$.  The merged Fermi surface within the first Brillouin zone is shown in Fig.~\ref{Fig-FS}. The Fermi surface consists of $4$ FS sheets: one quasi-two-dimensional tubular sheet and two nested ellipsoidal sheets very similar to OsB$_2$ \cite{Hebbache2009}.  An additional small $4^{\rm th}$ sheet nested inside the tubular sheet is also found for RuB$_2$.
 
\noindent
\emph{Summary and Discussion:}    
Using electrical resistivity $\rho(T)$, magnetic susceptibility $\chi(T)$, magnetization $M(H)$, and specific heat $C(T,H)$ data we have confirmed bulk superconductivity in RuB$_2$ with a superconducting critical temperature $T_c = 1.5$~K\@.  The $T = 0$ critical field is estimated to be $H_c(0) = 122$--$153$~Oe.  The magnitude of the anomaly in specific heat at $T_c$ in zero field is observed to be $\Delta C/\gamma _sT_c \approx 1.1$, which is much smaller than the value $1.43$ expected for a single-gap BCS superconductor.  This observation is similar to what has previously been observed for MgB$_2$ and OsB$_2$, and suggests multi-gap superconductivity in RuB$_2$.  This is confirmed by the excellent fitting of the electronic specific heat below $T_c$ to a two-gap model with the value of the two gaps estimated as $\Delta_1/k_BT_c \approx 1.88$ and $\Delta_2/k_BT_c \approx 1.13$.  The value of $\Delta C/\gamma T_c$ in a magnetic field becomes larger than its zero field value strongly indicating Type-I behaviour.  This is also similar to what was observed earlier for OsB$_2$ and also for other candidate Type-I superconductors like ScGa$_3$ and LaGa$_3$ \cite{Svanidze2012} and YbSb$_2$ \cite{Zhao2012}.  The $M(H_{eff})$ behaviour are also consistent with Type-I superconductivity.  This is confirmed by estimates of the Ginzburg-Landau parameter $\kappa$ which comes out to be $\kappa \approx 0.1 $--$0.6 < 0.707$.  These results strongly suggest that RuB$_2$ is a rare alloy Type-I superconductor and may be the first multi-gap Type-I superconductor.  We note that both YbSb$_2$ ($\kappa \approx 0.05$ and $\Delta C/\gamma T_c <$~BCS)\cite{Zhao2012} and boron-doped SiC ($\kappa \approx 0.35$ and $\Delta C/\gamma T_c <$~BCS) \cite{Kriener2008} have been reported as Type-I superconductors and have specific heat anomalies smaller than expected for single band BCS superconductivity.  However, both reported materials were multi-phase samples and in YbSb$_2$, an additional superconducting phase with a lower $T_c$ than the bulk $T_c$ was also observed, making it complicated to estimate intrinsic superconducting parameters.   Thus RuB$_2$ seems to be the best candidate for two-gap Type-I superconductivity so far.

 However, a scenario (anisotropic Type-I superconductivity) like the one recently suggested for OsB$_2$~\cite{Bekaert2016} could also be at play in RuB$_2$ and future work like imaging of magnetic flux entering the material may be useful to confirm the type of superconductivity in RuB$_2$.\\         
         
\noindent
\emph{Acknowledgments.--} We thank the X-ray facility at IISER Mohali.  JS acknowledges UGC-CSIR India for a fellowship.  DS thanks DST, India for INSPIRE faculty award (DST/INSPIRE/04/2015/000579).  YS acknowledges DST, India for support through Ramanujan Grant \#SR/S2/RJN-76/2010 and through DST grant \#SB/S2/CMP-001/2013.

\end{document}